# SOLVING CONTAINER TERMINALS PROBLEMS USING COMPUTER-BASED MODELING

**GAMAL ABD EL-NASSER A. SAID[1], ABEER M. MAHMOUD [2], EL-SAYED M. EL-HORBATY[3]**

Computer science Department, Faculty of Computer & Information Sciences, Ain Shams University, Egypt.

## ABSTRACT

This paper addresses the applications of different techniques for solving container terminals problems. We have built a simulation model that can be used to analyze the performance of container terminal operations. The proposed approach is intended to be applied for a real case study in Alexandria Container Terminal (ACT) at Alexandria port. The implementation of our approach shows that a proposed model increases the efficiency of Alexandria container terminal at Alexandria port.

## INTRODUCTION

The rising competition among ports has put considerable pressure on these ports to improve their performance. In most container terminals, a large portion of the container terminal turnaround time is spent on discharging and loading containers for a ship. In container terminals, quay cranes, yard cranes and trucks are used in different parts of a container terminal to transfer or transport containers from one location to another. Quay cranes handle containers by running at the berth to transfer containers between ships and trucks. Yard cranes are move in the yard to transfer containers between trucks and the container yard for container storage or retrieval. Trucks are highly mobile devices for transporting containers from one location to another [23].

The flow of containers in terminal operations is presented in Figure 1. Container terminals at seaports are the container shipping hubs of land-sea intermodal transportation. In container terminal, different types of handling equipment are used to transship containers from ships to storage at yard zone and vice versa. Container terminals are facing a set of interrelated NP-hard problems, crane assignment, truck assignment and storage space allocation problems.

In general terms, container terminal can be described as open systems of material flow with two external interfaces which are the quayside with loading and unloading of ships, and the landside where containers are loaded and unloaded on/off trucks. Containers are stored in stacks thus facilitating the decoupling of quayside and landside operation [13,19].

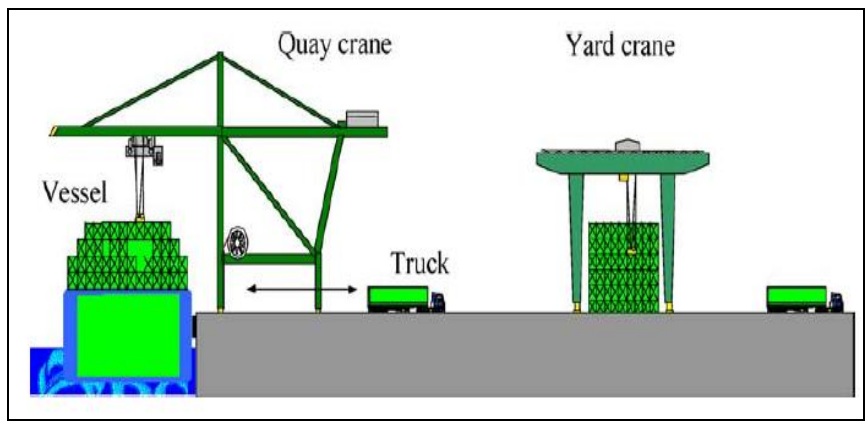

Figure 1 Typical flow of containers in terminal operations (Ng, 2005)

Thousand of containers are handled in a container terminal everyday by different types of material handling equipment. Managing activities of such high intensity level in a container terminal is a challenging task. In a container terminal, the allocation of resources is typically triggered by time, and all resources have to be considered simultaneously in the terminal's resource allocation process. Therefore many researchers try to find or improve different methods to solve these problems with high quality and in less time [17].

The operations of a container terminal are very complex which involve highly dynamic interactions between the units of various handling, transportation and storage. Container handling problems at container terminals are NP-hard, stochastic, nonlinear, and combinatorial optimization problems, thus requiring the application of heuristic algorithms to reach solutions [11,21,3]. Met-heuristics algorithms have been used to solve optimization problems, Among all of the heuristic algorithms such as : genetic Algorithm, tabu Search, and simulated annealing, genetic algorithms (GAs) are in wide application because of their ability to locate the optimal solution in the global solution space [1,7,8,9,10].

Cordeau et al. (2007) study the Service Allocation Problem, the objective is the minimization of container handling operations in the yard and it is formulated as a Quadratic Assignment Problem [16]. Gamal Abd El-Nasser et al. (2014) presents a comparative study between Meta-heuristic algorithms: Genetic Algorithm, Tabu Search, and Simulated Annealing for solving Quadratic Assignment Problem. The computational results show that genetic algorithm has a better solution quality than the other Meta-heuristic algorithms for solving Quadratic Assignment Problems [25]. Container handling problems at container terminal are too complex to be modeled analytically; discrete event simulation has been a useful tool for evaluating the performance of systems. However, simulation can only evaluate a given design, not providing optimization of such systems [14].

The rest of this paper is organized as follows: Section II presents the literature review for solving container terminal problems. Section III presents briefly discrete event simulation. Section IV discusses the methodology and framework developed to evaluate the performance of container terminal including the experimental results. Our conclusions and future work are given in section V.

## LITERATURE REVIEW

When a ship arrives at the port, multiple quay cranes are assigned for discharging the import containers from ships to trucks. Trucks transport discharged containers to storage yards. Yard Cranes are assigned for storing containers at storage yards for a certain period, there are different types as well as different sizes of containers; the type and the size of container have an effect on allocation of containers to the storage blocks.

Simulation modeling techniques are being applied to a wide range of container terminal planning processes and operational analysis of container handling systems. Discrete event simulation has long been a useful tool for evaluating the performance of complex systems [4]. Discrete-event simulation exploited to support container terminal decisions in a complex and stochastic environment. Simulation-based approaches have been widely used to model various planning problems arising in container terminals [15].

Jinxin et al.(2008) considered the problem of scheduling of trucks in a container terminal to minimize makespan. An integer programming model for truck scheduling and storage allocation problem is formulated. The problem was solved with a genetic algorithm [12]. Huang et al. (2008) presents a simulation model that can be used to simulate the container terminal operations for the purpose of terminal design, capacity planning and operations planning, it was found to be effective in replicating real-world operations as well as in evaluating the handling capacities [22].

K.L. Mak and D. Sun (2009) presents a new hybrid optimization algorithm combining the techniques of genetic algorithm and tabu search method to solve the problem of scheduling yard cranes to perform a given set of loading and unloading jobs in a yard zone [5]. Mohammad. B. et al. (2009) solved an extended Storage Space Allocation Problem (SSAP) in a container terminal by GA. The objective of the SSAP developed is to minimize the time of storage and retrieval time of containers [20]. Zeng and Yang (2009) presents a framework of simulation optimization. A simulation optimization model for scheduling loading operations in container terminals is developed to find good container loading sequences which are improved by a genetic algorithm through an evaluation process by simulation model to evaluate objective function of a given scheduling scheme [14].

I. Ayachi et al. (2010) presents a genetic algorithm (GA) to solve the container storage problem in the port. This problem is studied with different container types such as regular, open side, tank, empty and refrigerated containers. The objective of this problem is to determine an optimal containers arrangement, which respects customers' delivery deadlines, reduces the re-handle operations of containers and minimizes the stop time of the container ship [6].

Skinner et al. (2012) presents a modified mathematical model incorporates QC related operations. GA-based approach was presented to solve the job scheduling problem. The proposed approach has been fully implemented on a trial basis in the scheduling system and it effectively improves the performance of the container terminal [2]. Legat et al. (2012) present queuing-based representation of the current housekeeping process in a real container terminal and solve it by discrete-event simulation to i) assess the efficiency of the housekeeping operations under unforeseen events or process disturbances and ii) estimate the related productivity and waiting phenomena which, in turn, affect the vessel turn-around time [17].

Sriphrabu (2013) proposes a developed simulation model for stacking containers in a container terminal through developing and applying a genetic algorithm (GA) for containers location assignment with minimized total lifting time and increased service efficiency of the container terminals [18].

Table .1 shows the existing methods for solving container terminal problems. In this paper we have built a simulation model that can be used to simulate the container terminal operations and analyze the performance of Alexandria container terminal.

Table .1 Existing methods for solving container terminal problems

| Ref. no. | Technique Used | Problem solved | Weakness |
|---|---|---|---|
| 12 | Integer programming | Truck scheduling and storage allocation problem | Problems solved separately As such, they are unable to solve the container terminal problems as a whole |
| 5 | genetic algorithm and tabu search | Yard cranes scheduling | |
| 6,20 | genetic algorithm | Storage Space Allocation Problem | |
| 15 | Discrete-event simulation | Yard cranes | |
| 17 | Discrete-event simulation | Housekeeping process | |
| 14 | Simulation based optimization | scheduling loading operations | |
| 18 | Simulation based optimization | stacking containers | |

**DISCRETE EVENT SIMULATION**

Modeling is the process of producing a model; a model is a representation of the construction and working of some system of interest. A model is similar to but simpler than the system it represents. It enables us to study the impact from changes and reduce the risk that might occur in a real situation. It also allows us to measure the capabilities of different alternatives of the system being designed. One purpose of a model is to enable the analyst to predict the effect of changes to the system. On the other hand, it should not be so complex that it is impossible to understand and experiment with it.

Simulation is a tool to evaluate the performance of a system, under different configurations of interest and over long periods of real time. In discrete event simulation the central assumption of the system changes instantaneously in response to certain discrete events. The simulation method alone can only provide feasible solutions of certain conditions of the systems but not the optimal solutions [4].

**METHODOLOGY AND FRAMEWORK**

The container handling process at a container terminal including the most commonly employed facilities: quay cranes, yard cranes, trucks and storage. The container terminal problems are known to be NP-hard and the computation complexity increases exponentially. When a vessel carrying containers arrives at a container terminal, multiple quay cranes are assigned for discharging jobs. Trucks move to the quayside to transport the unloaded containers to the yard side for storage. In each yard zone, assigned yard cranes discharge containers from Trucks to storage. When the discharging job finishes, the trucks, yard cranes and quay cranes are released and readied for assignment to their next job.

We have built a simulation model that can be used to simulate container terminal operation. The objective of the model is to analyze the performance of container terminal operations for increasing throughput of container terminal. Computational experiments were conducted to analyze the performance of container terminal operation. The proposed approach is intended to be applied for a real case study in Alexandria Container Terminal (ACT) at Alexandria port.

**Model Validation and verification**

For purposes of validation of simulation model and verification of simulation flexsim program, the results of simulation model were compared with the actual measurement. Several statistics were used as a comparison between simulation output and real data: annual throughput of a berth, berth occupancy rate, ship service time and the total time that ship spends in port, handling units per hour per ship, average number of assigned Quay cranes, yard cranes utilization, storage blocks at yards, the number of ship berthing and container terminal throughput.

**Case Study - Alexandria Container Terminal (ACT)**

Container Terminal (CT) plan is important for allocating resources and its usage within the CT such as quay area, yard areas with its different usages export, import, reefers, Hazardous. The present Throughput of (ACT) at Alexandria port is 500000 TEU (twenty-foot equivalent unit) per Year. Fig. 2 shows ACT layout plan, demonstrates the main Alexandria container terminal operation areas at Alexandria port.

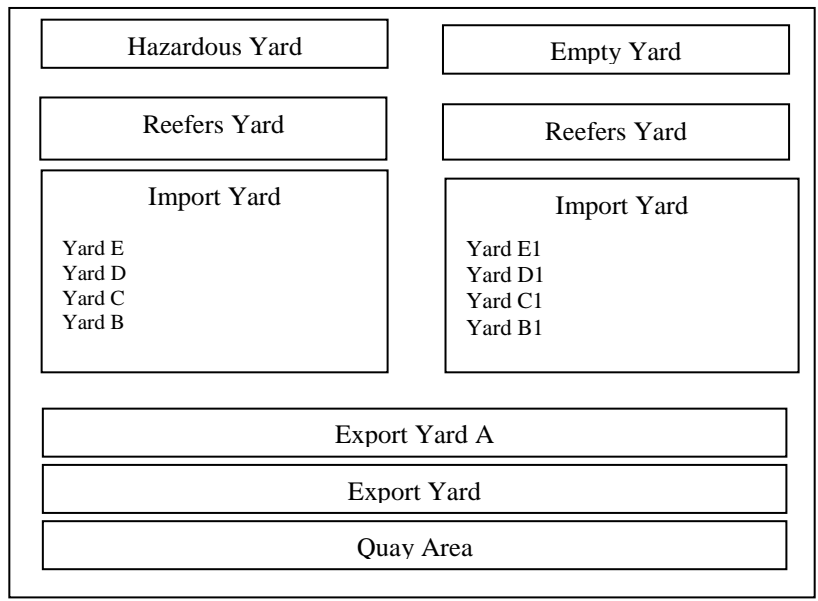

Figure 2 The Layout plan of container terminal (ACT)

The scope of our experiment model is as follows:

| | |
|---|---|
| Container quay length | 530 meter |
| Water Depth | 14 meter |
| Terminal Area | 163000 m$^2$ |
| Number of Quay cranes | 5 |
| Number of Yard Cranes | 8 |
| Number of Trucks | 25 |
| Heavy Top Lift Truck | 15 |
| Empty Handler Side Spreader | 8 |

Implementation of the model was run on a Laptop with the following configurations: i3 CPU 2.4 GHZ, 4.0 GB RAM, Windows 7 using discrete event simulation software (Flexsim). A distinct advantage of Flexsim software over similar software like Arena is that it comes with flexsim CT, a library specifically designed for simulating container terminal operation. The model overall structure of our experiment model is presented in Figure 3, and a snapshot from the animation is depicted in Figure 4.

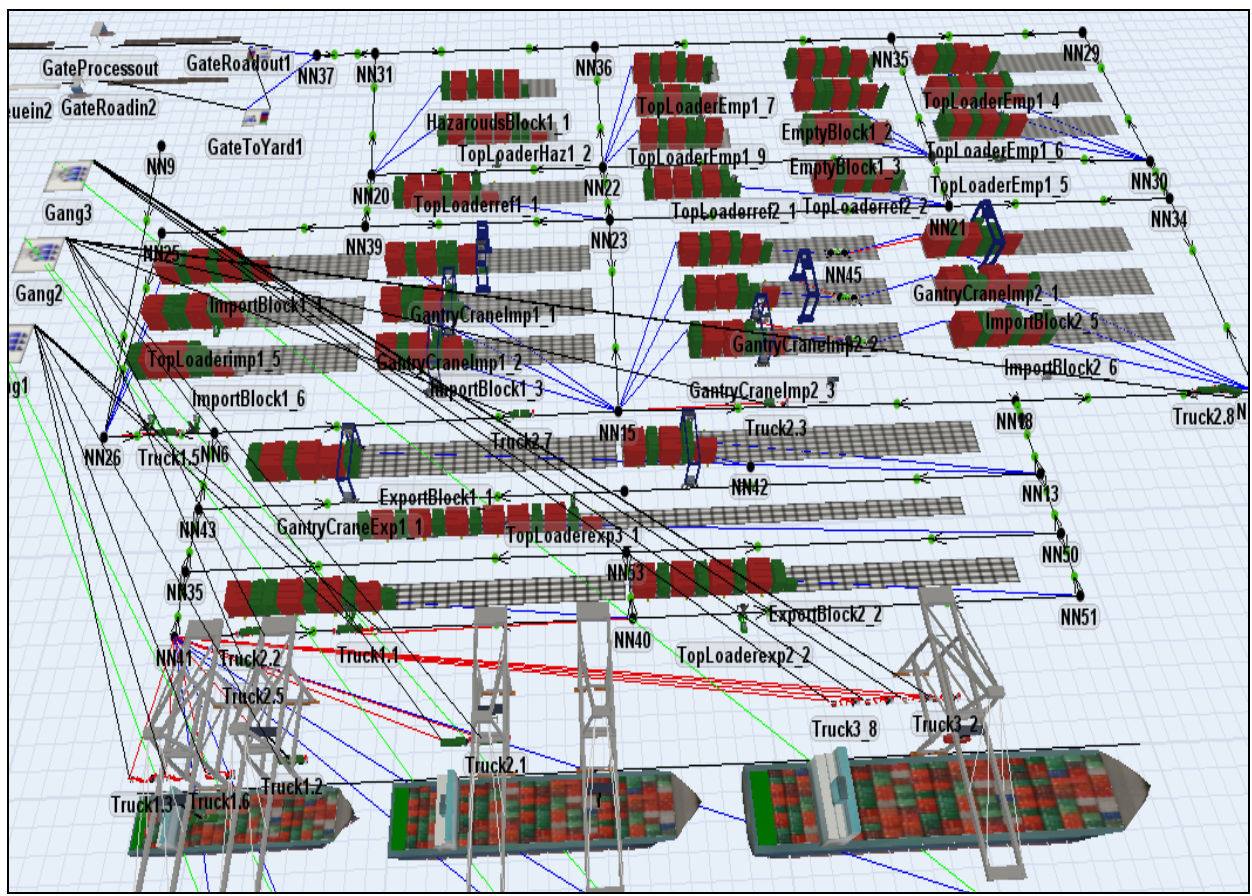

Figure 3 Container Terminal simulation model

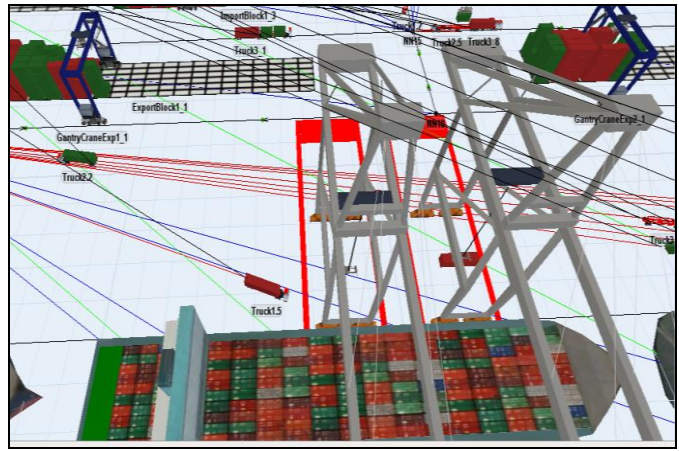

Figure 4 A snapshot from the animation

We simulated this model which represents container terminal operation for one week. The container types are 20 and 40 foot containers. The results show that a proposed model increases the efficiency of containers handling at Alexandria container terminal in Alexandria port form 500000 TEU per year to 730000 TEU per Year with a minimum use of different expensive equipment. The berth planner for one week is presented in Figure 5 shows that 14000 TEU can be loaded or discharged per week. The horizontal axis represents the berth and the length of the berth is displayed at the top of the graph. The vertical axis represents time beginning on Monday and ending on Sunday for each week in the schedule.

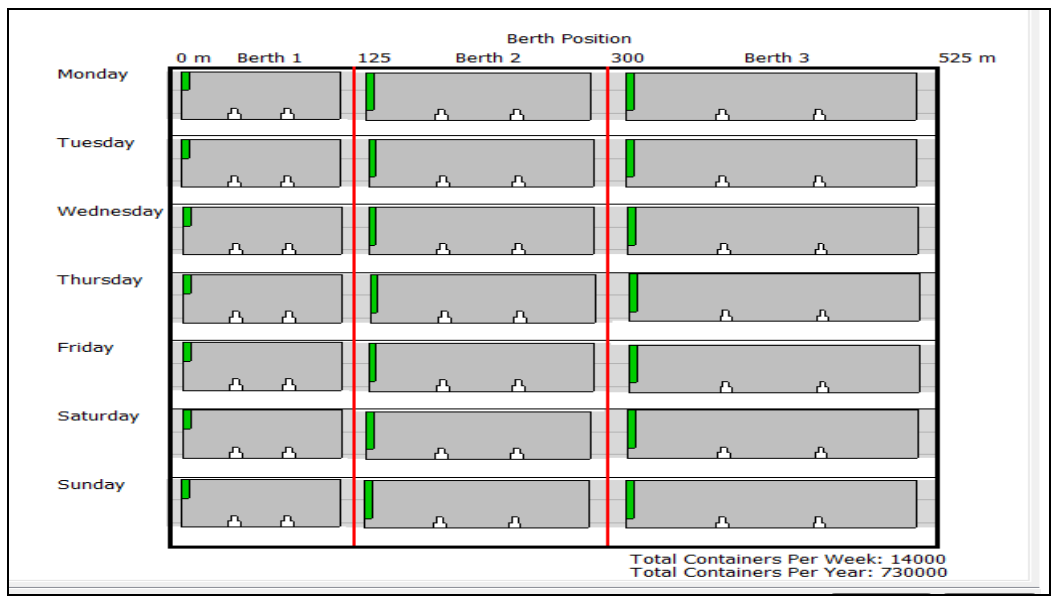

Figure 5 Berth planner

When a crane is working on a ship it should be busy. However, if there are not enough trucks to service the crane then the crane will be idle part of the time. Table 2 shows the percentage of working, net moves per hour, and throughput of each quay crane. Results show that the Net Moves per Hour of Crane 5 more than other cranes because crane 5 was assigned for long container ships.

Table .2 Quay cranes Statistics

| Quay Cranes | Working % | Net Moves Per Hour | Throughput |
|---|---|---|---|
| Crane 1 | 40.3 | 24.8 | 2771 |
| Crane 2 | 38.7 | 25.5 | 2735 |
| Crane 3 | 39 | 24.4 | 2644 |
| Crane 4 | 41.1 | 24.8 | 2827 |
| Crane 5 | 38.4 | 32.3 | 3444 |
| Total | | | 14421 |

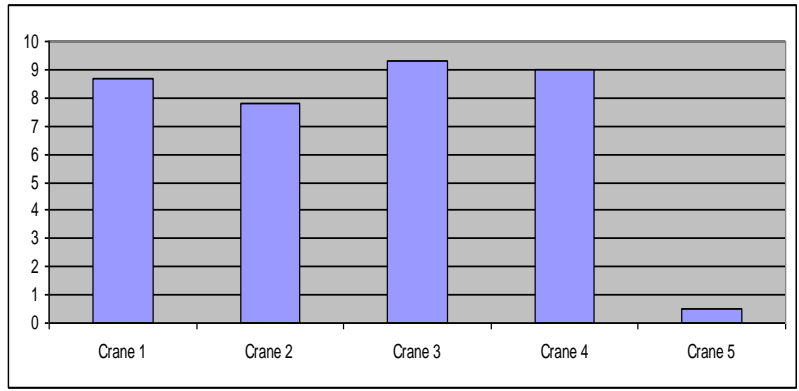

Figure 6 Quay Cranes waiting for Trucks

Figure 6 illustrates quay cranes waiting time for trucks. The X-axis represents quay cranes; The Y-axis represents the average waiting time of each crane for trucks. Crane 5 less waiting time than other cranes.

Table 3 shows the percentage of working and net moves per hour of each RTG (Rubber Tyred Gantry crane), heavy top Lift & empty handler spreader at (Import, Export, Empty, Reefers, Hazardous) Yards. Results show that the percentage of working of RTG cranes at import yards more than other yards because the number of containers at import yards more than other yards.

Table .3 Yard Cranes Statistics

| Yard Cranes | Working % | Net Moves Per Hour |
|---|---|---|
| Gantry Crane Export – Yard B1 | 21 | 14.5 |
| Gantry Crane Export - Yard B2 | 18.9 | 16 |
| Gantry Crane Import - Yard C1 | 30.5 | 21 |
| Gantry Crane Import - Yard C2 | 32.2 | 19.1 |
| Gantry Crane Import - Yard D1 | 28.5 | 21.1 |
| Gantry Crane Import - Yard D2 | 33.4 | 18.5 |
| Gantry Crane Import - Yard E1 | 29.5 | 21.3 |
| Gantry Crane Import - Yard E2 | 34.7 | 18.4 |
| Heavy Top Lift & Empty Handler Spreader – (Empty, Export A, Reefers, Hazardous) Yards | 21 | 15.5 |

Table 4 shows the average number of containers at (Import, Export, Empty, Reefers, Hazardous) Yards. Results show that the number of containers at import yards more than the other yards.

Table 4 Yard Blocks Transactions

| Storage Blocks | Average Transactions |
|---|---|
| Import containers | 2046 |
| Export containers | 1040 |
| Empty containers | 1548 |
| Reefer containers | 543 |
| Hazardous containers | 369 |

Table 5 shows the number of containers transported by trucks.

Table .5 Trucks Statistics

| Trucks | Throughput |
|---|---|
| Terminal Tractors | 14421 |

**CONCLUSION AND FUTURE WORK**

In this paper we have built a simulation model that can be used to simulate the container terminal operations and analyze the performance of Alexandria Container terminal depends upon Quay cranes, Yard Cranes, Trucks, and Yard Blocks. The objective of the model is to analyze the performance of container terminal operations for improving the efficiency of Alexandria container terminal. Computational experiments were conducted to analyze the performance of container terminal operation. In case of ACT, the average container handling per year equals to 500000 TEU per year. However, the Obtained solution value equal to 730000 TEU per Year which is better throughput. This means that the model resulted in a better equipment utilization and storage space allocation than the methods that ACT planners use.

In the future, analytical and empirical evaluations will be done for solving container handling problems; crane assignment, truck assignment and storage space allocation problems using simulation based optimization technique and develop a computational framework for enhancing the computational efficiency of the proposed solution technique.

**REFERENCES**

1. Peng YANG. Junqing SUN, Wenying YUE, "A Novel Genetic Algorithm for Multiple Yard Cranes Scheduling", Journal of Computational Information Systems 9: 14 5761–5769, 2013.

2. Bradley Skinner a, Shuai Yuan a, Shoudong Huang a, Dikai Liu a, Binghuang Cai a, Gamini Dissanayake a, Haye Lau b, Andrew Bott b, Daniel Pagac, " Optimisation for job scheduling at automated container terminals using genetic algorithm", Computers & Industrial Engineering,DOI:10.1016, 09/2012.

3. Wenbin HU, Zhengbing HU, Lei SHI, Peng LUO and Wei SONG, "Combinatorial Optimization and Strategy for Ship Stowage and Loading Schedule of Container Terminal". Journal of computers, vol. 7, NO. 8, August 2012.

4. Albrecht, M., Introduction to Discrete Event Simulation, 2010.

5. K.L. Mak, D. Sun., "Scheduling Yard Cranes in a Container Terminal Using a New Genetic Approach", Engineering Letters, Vol. 17 Issue 4, p274 , 2009.

6. I. Ayachi, R. Kammarti,, M. Ksouri and P. Borne., "A Genetic algorithm to solve the container storage space allocation problem", EEE International conference on Computational Intelligence and Vehicular System; 09/2010.

7. H. Javanshir, S. R. Seyedalizadeh Ganji., "Yard crane scheduling in port container terminals using genetic algorithm", J. Ind. Eng. Int., 6 (11), 39-50, spring, ISSN: 1735-5702, 2010.

8. H. L. Ma, F. T. S. Chan, S. H. Chung and C. S. Wong., "Berth Allocation Planning for Improving Container Terminal Performance", Proceedings of the International Conference on Industrial Engineering and Operations Management Istanbul, Turkey, July 3 – 6,2012.

9. Pragya Gupta, SSCET, Bhilai., "Solving Container Loading Problem Using Improved Genetic Algorithm", International Journal of Engineering Research & Technology (IJERT), ISSN: 2278-0181,Vol. 1 Issue 8,2012.

10. Z.X. Wang, Felix T.S.Chan, and S.H. Chung., "Storage Allocation and Yard Trucks Scheduling in Container Terminals Using a Genetic Algorithm Approach", 3rd International Conference on Intelligent Computational Systems (ICICS'2013), Singapore, 2013.

11. Lee D.H., Wang H.Q., Miao L.X., "Quay crane scheduling with noninterference constraints in port container terminals." Transportation Research, Part E 44, 2008.

12. CAO Jinxin, SHI Qixin, Der-Horng Lee, "A Decision Support Method for Truck Scheduling and Storage Allocation Problem at Container", Tsinghua science and technology ISSN 1007-0214 34/67 pp211-216, Volume 13, Number S1, October 2008.

13. D. Steenken, S. VoB, and R. Stahlbock. "Container terminal operation and operations research - a classification and literature review". OR Spectrum, 26:3-49, 2004.

14. Q. Zeng and Z. Yang. "Integrating simulation and optimization to schedule loading operations in container terminals". Computers and Operations Research, 36:1935-1944, 2009.

15. X. Guo, S. Huang, W. Hsu, M. Low, "Yard crane dispatching based on real time data driven simulation for container terminals", in: S. Mason, R. Hill, L. Mnch, O. Rose, T. Jefferson, J. Fowler (Eds.), Proceeding of the Winter Simulation Conference, pp. 2648–2655, 2008.

16. Cordeau, J. F., M. Gaudioso, G. Laporte and L. Moccia, "The service allocation problem at the Gioia Tauro maritime terminal", European Journal of Operational Research, 176 (2) 1167–1184, 2007.

17. Pasquale Legat, Rina Mary Mazza, Roberto Trunfio, "Simulation for performance evaluation of the housekeeping process". Proceedings of the 2012 Winter Simulation Conference, 2012.

18. Phatchara Sriphrabu, Kanchana Sethanan, and Banchar Arnonkijpanich., "A Solution of the Container Stacking Problem by Genetic Algorithm ". IACSIT International Journal of Engineering and Technology, Vol. 5, No. 1, February 2013.

19. Chuqian Zhang, Jiyin Liu, Yat wah Wan, Katta G. Murty, and Richard J. Linn, "Storage space allocation in container terminals". Transportation Research Part B 37: 883–903, 2003.

20. Mohammad. B, Nima. S, Nikbakhsh. J, "A genetic algorithm to solve the storage space allocation problem in a container terminal", Computers & Industrial Engineering 56: 44–52, (2009).

21. Jula H., Dessouky M., Ioannou P., Chassiakos A., "Container movement by trucks in metropolitan networks: modeling and optimization", Transportation Research Part E 41, pp. 235–259, 2005.

22. S. Y. Huang, W. J. Hsu, C. Chen, R. Ye and S. Nautiyal, "Capacity Analysis of Container Terminals Using Simulation Techniques", International Journal of Computer Applications in Technology, Vol. 32, No. 4, 2008.

23. Ng W C. Crane scheduling in container yards with inter-crane interference. European Journal of Operational Research, 2005, 164: 64-78.

24. Azmi Alazzam and Harold W. Lewis, "A New Optimization Algorithm For Combinatorial Problems", (IJARAI) International Journal of Advanced Research in Artificial Intelligence, Vol. 2, No.5, 2013.

25. Gamal Abd El-Nasser, Abeer M. Mahmoud and El-Sayed M. El-Horbaty, "A Comparative Study of Meta-heuristic Algorithms for Solving Quadratic Assignment Problem", (IJACSA) International Journal of Advanced Computer Science and Applications, ISSN: 2156-5570, Vol. 5, No.1, 2014.

26. Flexsim software products inc.: www.flexsim.com.

27. Alexandria Container terminal: www.alexcont.com [accessed 25/2/2014]

## AUTHOR'S PROFILE

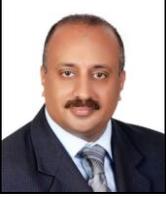

**Gamal Abd El-Nasser A. Said**: He received his M.Sc. (2012) ) in computer science from College of Computing & Information Technology, Arab Academy for Science and Technology and Maritime Transport (AASTMT), Egypt and B.Sc (1990) from Faculty of Electronic Engineering, Menofia University, Egypt.

His work experience as a Researcher, Maritime Researches & Consultancies Center, Egypt. Computer Teacher, College of Technology Kingdom Of Saudi Arabia and Lecturer, Port Training Institute, (AASTMT), Egypt. Now he is Ph.D. student in computer science, Ain Shams University. His research areas include optimization, discrete-event simulation, and artificial intelligence.

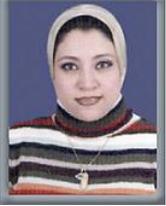

**Dr Abeer M. Mahmoud**: She received her Ph.D. (2010) in Computer science from Niigata University, Japan, her M.Sc (2004) B.Sc. (2000) in computer science from Ain Shams University, Egypt.

Her work experience is as a lecturer assistant and assistant professor, faculty, of computer and information sciences, Ain. Shams University. Her research areas include artificial intelligence medical data mining, machine learning, and robotic simulation systems.

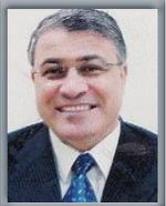

**Professor El-Sayed M. El-Horbaty**: He received his Ph.D. in Computer science from London University, U.K., his M.Sc. (1978) and B.Sc (1974) in Mathematics From Ain Shams University, Egypt. His work experience includes 39 years as an in Egypt (Ain Shams University), Qatar (Qatar University) and Emirates (Emirates University, Ajman University and ADU University). He Worked as Deputy Dean of the faculty of IT, Ajman University (2002-2008). He is working as a Vice Dean of the faculty of Computer & Information Sciences, Ain Shams University (2010-Now). Prof. El-Horbaty is current areas of research are parallel algorithms, combinatorial optimization, image processing. His work appeared in journals such as Parallel Computing, International journal of Computers and Applications (IJCA), Applied Mathematics and Computation, and International Review on Computers and software.